\documentclass[uslettersize, 10pt, conference]{ieeeconf}      

\IEEEoverridecommandlockouts                              
\usepackage{graphics} 
\usepackage{epsfig} 
\usepackage{mathptmx} 
\usepackage{times} 
\usepackage{amsmath} 
\usepackage{amssymb}  
\usepackage{amscd}
\usepackage{subcaption}
\usepackage{cite}
\usepackage{float}
\usepackage[export]{adjustbox}

\newtheorem{theorem}{Theorem}
\newtheorem{lem}[theorem]{Lemma}

\title{\LARGE \bf
Decentralized Formation Control with A Quadratic Lyapunov Function}

\author{Xudong Chen$^{1}$ 
\thanks{This work has been partially supported by grant NSF ECCS 13-51586.}
\thanks{$^{1}$Xudong Chen is with Department of Electrical and Computer Engineering and Coordinated Science Laboratory, University of Illinois at Urbana-Champaign, email:
        {\tt\small xdchen@illinois.edu}}%
}

\begin{document}

\maketitle
\thispagestyle{empty}
\pagestyle{empty}

\begin{abstract} In this paper, we investigate a decentralized formation control algorithm for an undirected formation control model. Unlike other formation control problems where only the shape of a configuration counts, we emphasize here also its Euclidean embedding.  By following this decentralized formation control law, the agents  will converge to certain equilibrium of the control system. In particular, we show that there is   a quadratic Lyapunov function associated with the formation control system whose unique local (global) minimum point is the target configuration. In view of the fact that there exist multiple equilibria (in fact, a continuum of equilibria) of the formation control system, and hence there are solutions of the system which converge to some equilibria other than the target configuration, we apply simulated annealing, as a heuristic method, to the formation control law to fix this problem. Simulation results show that sample paths of the modified stochastic system approach the target configuration. 
\end{abstract}


\section{Introduction}
Undirected formation control has been one of the most studied subjects in multi-agent systems. The formation control model is described by two characteristics: one is an undirected graph describing the pattern of interaction, and the other is a set of scalar functions each of which describes the interaction law between a pair of adjacent agents. A detailed description is given below
\vspace{3pt}

Let $\mathbb{G} = (V,E)$ be an undirected connected graph of $N$ vertices $V=\{1,\cdots,N\}$. Let $V_i:= \{j\in V|(i,j)\in E\}$ be the set of vertices adjacent to vertex $i$. We then consider the motion of $N$ agents $\vec a_1,\cdots,\vec a_N$ in $\mathbb{R}^n$ by      
\begin{equation}\label{MODEL}
\frac{d}{dt} {\vec a}_i = \sum_{j\in V_i} u_{ij}\cdot (\vec a_i-\vec a_j), \hspace{10pt} \forall i = 1,\cdots,N
\end{equation}
Each $u_{ij}$ with $(i,j)\in E$ is a scalar function describing how adjacent agents $\vec a_i$ and $\vec a_j$ interact with each other. We require that $u_{ij}$ be identical with $u_{ji}$ for all $(i,j)\in E$, in other words, interactions among agents are reciprocal.  In its most general form, each $u_{ij}$ could be a function of $\vec a_1,\cdots,\vec a_N$, and possibly the time variable $t$ as well. In this case, we have proved in  \cite{CDC2014}  that if $\mathbb{G}$ is connected, then system \eqref{MODEL}, treated as a centralized control system, is approximately path-controllable over an open dense subset of the configuration space. Yet, if we regard system \eqref{MODEL} as a decentralized control system, i.e, each agent $\vec a_i$ only accesses part of the information, then there is a restriction on what variables each $u_{ij}$ can depend on. For example,  it is often assumed that each agent $\vec a_i$ knows agent $\vec a_j$ if and only if $\vec a_j$ is its neighbor, i.e, $j\in V_i$. Then in this case each $u_{ij}$ is at most a function of $\vec a_i$ and $\vec a_j$, and possibly the time $t$. 
\vspace{3pt}

Over the last decade, there have been many solid works about using system \eqref{MODEL} to achieve decentralized formation control. Questions about the level of interaction laws that are necessary for organizing such systems, questions about system convergence, questions about counting and locating stable equilibria, and questions about the issues of robustness and etc. have all been treated to some degree (see, for example,  \cite{GP,KMB,AB3,AH,A3,USZB,ZSB,SAB,OR2,GFOM,CDC2014,CDC2015}).  For example, a popular decentralized algorithm,  known as the Krick's law \cite{KMB,AB3,AH,A3,USZB,ZSB,SAB}, is that we assume each agent $\vec a_i$ measures the mutual distance between himself and its neighbors $\vec a_j$. The control law is then given by $u_{ij} = |\vec a_i-\vec a_j|^2 - d^{*2}_{ij}$ for all $(i,j)\in E$ where $d^*_{ij}$  is the prescribed distance between $\vec a_i$ and $\vec a_j$ in the target formation. By following this decentralized algorithm, system \eqref{MODEL} will then be a gradient system with respect to the potential function $\Phi(\vec a_1,\cdots,\vec a_N) = \sum_{(i,j)\in E}(|\vec a_i-\vec a_j|^2 - d^{*2}_{ij})^2/4$. In fact, it has been shown that  if each $u_{ij}$ is a continuous function depending only on the distance $|\vec a_i-\vec a_j|$, then the resulting system is always a gradient system \cite{GP,GFOM}. 
However, the associated potential function $\Phi$ often has multiple local minima, and in some cases, the number of local minima has an exponential growth with respect to the number of agents (see, for example, \cite{A3,CDC2015}).  
\vspace{3pt}

Also, we note that if each $u_{ij}$ depends only on the mutual distance $|\vec a_i-\vec a_j|$ as is the case if we adopt the Krick's law for $u_{ij}$, then  how a configuration is embedded into the Euclidean space is not relevant, only the shape of the formation matters. In other words, if a configuration is an equilibrium associated with system \eqref{MODEL}, then any rotation or translation of the configuration will also be an equilibrium. In any of such  case, the group  of rigid motions is introduced to describe this phenomena. Two configurations will be recognized as the same target formation if they are in the same orbit with respect to the group action.       
\vspace{3pt}

In this paper, as in many earlier work on this problem, we will investigate system \eqref{MODEL}, treated as a decentralized formation control system, by equipping it with a set of new control laws. What distinguishes this paper from others is that in addition to the shape of the target configuration, we also emphasize its Euclidean embedding. To be more precise, we let $p = (\vec a_1,\cdots, \vec a_N)$ and $p' = (\vec a'_1,\cdots, \vec a'_N)$ be two configurations with the same centroid, i.e, $\sum^N_{i=1} a_i = \sum^N_{i=1} a'_i$, and we distinguish $p$ and $p'$ in the sense that these two configurations are recognized as the same target formation  if and only if $p = p'$. We impose the condition that  $p$ and $p'$ have the same centroid  because of the fact that the centroid of a configuration in an undirected formation control system is invariant along the evolution regardless of the choice of the control laws.  
\vspace{3pt}

The decentralized control law is then designed for each agent so that  the solution of the control system may converge to the target configuration.  In particular, we show that   there is a quadratic Lyapunov function associated with system \eqref{MODEL} whose unique local (global) minimum point is the target configuration. But we also note (and we will see later in the paper) that  there may exist a continuum of equilibria for system \eqref{MODEL}, thus a solution of system \eqref{MODEL} may fail to converge to the global minimum point. To fix this problem, we then modify the formation control laws by adding  noise terms. This is an application of simulated annealing to formation control systems. Simulation results then show that  sample paths of the modified stochastic system approach the global minimum point. 
\vspace{3pt}

The rest of this paper is organized as follows. In section 2, we will first specify what information each agent knows, or in other words what variables $u_{ij}$ can depend on. Then we will introduce the decentralized formation  control law, and establish the convergence of system \eqref{MODEL}. In section 3, we will explore one of the limitations of this formation control model by showing that there may exist a continuum of equilibria of system \eqref{MODEL}. For simplicity,  we will only focus on trees as a special type of network topologies. In this special case,  we show that there is a simple condition for determining whether a configuration $p$ is an equilibrium or not, and thus there is a geometric characterization  
of the set of equilibria of system \eqref{MODEL}. The existence of continuum of equilibria poses a problem about the convergence of system \eqref{MODEL} to the target configuration. 
In section 4, we will focus on fixing this problem by applying the technique of simulated annealing to the algorithm.  Simulation results  then show that a typical sample path will converge to the target configuration.

\section{A Lyapunov Approach for \\Decentralized Formation Control}
In this section, we will introduce the decentralized formation control law and show that system \eqref{MODEL}, when equipped with this control law,  converges to the set of equilibria. But before that, we need to be clear about what we mean by a decentralized formation control system. So in the first part of this section, 
we will specify what information each agent knows, i.e, what variables each $u_{ij}$ can depend on. Also we will specify how the information of the target configuration is distributed among agents. 

\subsection{Information distribution among the agents}
We first introduce the underlying space of system \eqref{MODEL}. As interactions among agents are reciprocal, the centroid of a configuration is always invariant along the evolution in an undirected formation control system, so we may as well assume that the centroid of a configuration is located at the origin. The configuration space $P$, as the underlying  space of system \eqref{MODEL}, is then defined by
\begin{equation}
P := \big \{p=(\vec a_1,\cdots,\vec a_N)\in \mathbb{R}^{n\times N}\big | \sum^N_{i=1}\vec a_i=0\big \}
\end{equation} 
It is clear that $P$ is a Euclidean space of dimension $n\times (N-1)$.   In this paper, we assume that 
\begin{equation}
q=(\vec b_1,\cdots,\vec b_N)\in P
\end{equation}
is the target configuration, i.e, each $\vec b_i$ is the target position for agent $\vec a_i$. We will now specify what information each agent $\vec a_i$ can access.  In this paper, if $(i, j)$ is an edge of $\mathbb{G}$, we then assume that 
\vspace{2pt}
\begin{itemize}
\item[a)] agent $\vec a_i$ knows $(\vec b_j-\vec b_i)$;
\vspace{2pt}
\item[b)] agent $\vec a_i$ is able to measure $(\vec a_j(t)-\vec a_i(t))$ at any time $t$;
\end{itemize}
\vspace{2pt}
Consequently, if $(i, j)$ is an edge of $\mathbb{G}$, then we require that each scalar function $u_{ij}$ depend only on $(\vec a_j(t)-\vec a_i(t))$, $(\vec b_j-\vec b_i)$ and possibly the time variable $t$. 

\subsection{The decentralized formation control law}

Suppose $(i, j)$ is an edge of $\mathbb{G}$, we then let the control law $u_{ij}$ be defined as
\begin{equation}\label{DEFIU}
u_{ij}:=\left\{
\begin{array}{ll}
\langle \vec b_j - \vec b_i, \vec a_j - \vec a_i \rangle / |\vec a_j-\vec a_i|^2-1 & \vec a_j \neq \vec a_i\vspace{3pt}\\
0 &  \vec a_j = \vec a_i
\end{array}\right.
\end{equation}
where $\langle\cdot,\cdot\rangle$ is the standard inner-product of two vectors, and $|\cdot|$ is the standard Euclidean norm of a vector. We note that in this definition, if we exchange roles of vertex $i$ and vertex $j$, then $u_{ji}$ will be identical with $u_{ij}$. This is consistent with our assumption that the formation control system is undirected. 
The main result of this section we will prove is about the convergence of system \eqref{MODEL} as stated below.
\vspace{5pt}


\begin{theorem}
Let $u_{ij}$ be the decentralized control law defined by expression \eqref{DEFIU}. Then there is a quadratic Lyapunov function associated with system \eqref{MODEL} defined as 
\begin{equation}\label{PSIEQ}
\Phi(\vec a_1,\cdots,\vec a_N):= \sum^N_{i=1} |\vec a_i-\vec b_i|^2
\end{equation}
Let $p(t) = (\vec a_i(t),\cdots,\vec a_N(t))$ be a solution of system \eqref{MODEL}, then 
\begin{equation}\label{GRLI}
\frac{d}{dt}\Phi(p(t)) = -\sum_{(i,j)\in E} u^{2}_{ij}\cdot |\vec a_i(t)-\vec a_j(t)|^2 \le 0
\end{equation}  
The derivative  is zero if and only if $p(t)$ is an equilibrium. 
\end{theorem}
\vspace{5pt}

\begin{proof} 
This proof is done by explicit computation. We check that  
\begin{equation}
\begin{array}{lll}
\displaystyle \frac{d}{dt}\Phi(p) & = & -\sum^N_{i=1}\big \langle \displaystyle\frac{d}{dt}{\vec a}_i, \vec b_i - \vec a_i \big \rangle \vspace{3pt}\\
\vspace{2pt}
& = & -\sum^N_{i=1} \big \langle \sum_{j\in V_i}u_{ij}\cdot (\vec a_i - \vec a_j), \vec b_i-\vec a_i\big  \rangle \vspace{3pt}\\  
\vspace{2pt}
& = & -\sum_{(i,j)\in E}  u_{ij}\cdot \big \langle (\vec a_i-\vec a_j), (\vec b_i-\vec b_j)-(\vec a_i-\vec a_j)\big   \rangle \vspace{3pt}\\
\vspace{2pt}
& = & -\sum_{(i,j)\in E}  u^2_{ij}\cdot  |\vec a_i-\vec a_j|^2   
\end{array}
\end{equation}
It is then clear that the time derivative is zero  if and only if each $u_{ij} \cdot (\vec a_i - \vec a_j)$ is zero which implies that $p$ is an equilibrium.   
\end{proof}
\vspace{5pt}
{\it Remark I}. There may exist multiple equilibria of system \eqref{MODEL}. In fact, as we will see in the next section that in the case $\mathbb{G}$ is a tree graph there exists a continuum of equilibria. Nevertheless, 
there is only one local (and also global) minimum point of the potential function $\Phi$ which is $q$. It thus suggests that we apply simulated annealing to this formation control law as we will discuss  in the last section of the paper.  
\vspace{5pt}
\\
{\it Remark II}. Notice that the potential function $\Phi(p)$ approaches  infinity as $|p-q|$ goes to infinity.  On the other hand, we have 
\begin{equation}
\Phi(p(t)) = |p(t) - q| \le |p(0) - q| = \Phi(p(0))
\end{equation}
So each solution $p(t)$ of system \eqref{MODEL} has to remain in a bounded set, and thus converges to the set of equilibria. In other words,  no agent escapes  to infinity along the evolution.

\section{Existence of Continuum of Equilibria}\label{TREE}
In this section, we will explore the set of equilibria of system \eqref{MODEL}. The main purpose of doing this is to illustrate one of the limitations of this formation control law. It is well-known that if $\Phi$ is a Lyapunov function for a dynamical system $\dot{x} = f(x)$ and $x_0$ is a unique equilibrium, then $x_0$ is stable and all solutions of the system will converge to $x_0$. However, this is not the case here, i.e, the target configuration $q$ will not be the unique equilibrium of system \eqref{MODEL}. As we will see in this section there may exist a continuum of equilibria of system \eqref{MODEL}. For simplicity, we will only focus on the case where the interaction pattern $\mathbb{G}$ is a tree graph. We focus on this special class of interaction patterns because in this case there is a simple condition telling us whether a configuration $p$ is an equilibrium or not.  In particular, we will use this condition to characterize the  set of equilibria of system \eqref{MODEL} in a geometric way.  
\vspace{3pt}

\begin{figure}[h]
\begin{center}
\includegraphics[scale=.4]{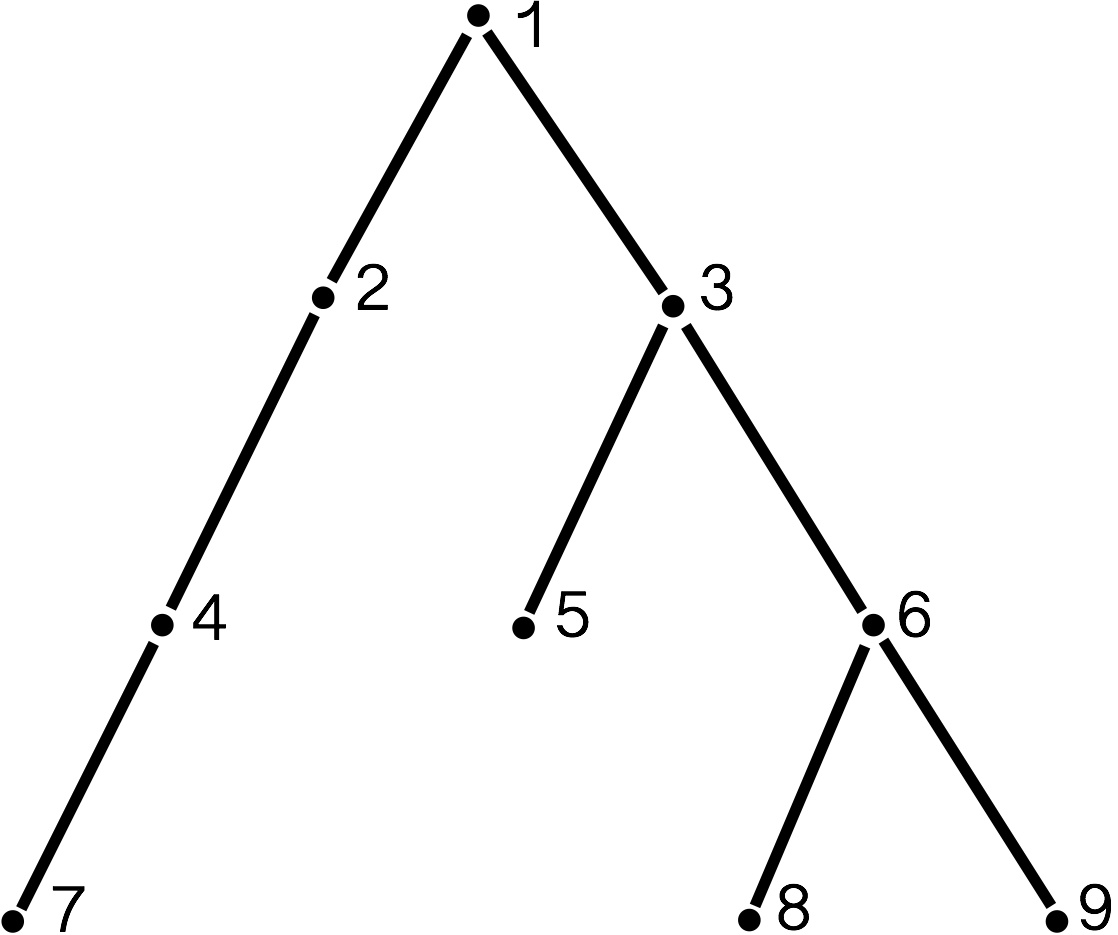}
\caption {A tree graph of $9$ vertices and $8$ edges. Vertices are labeled with respect to an inductive construction. For any two vertices in the graph, there is a unique path connecting them. The four vertices $5$, $7$, $8$ and $9$ are leaves of the tree graph above.}
\label{tree}
\end{center}
\end{figure}

A path in a graph $\mathbb{G}$ is a finite sequence of edges which connects a sequence of vertices. A simple path then refers to a path which does not have repeated vertices, and a circle refers to a path without repeated vertices or edges, other than the repetition of the starting and ending vertices. An undirected graph $\mathbb{G}$ is a {\bf tree} if any two vertices of $\mathbb{G}$ are connected by a unique simple path, i.e, there is no circle in $\mathbb{G}$. Each tree graph can be inductively built up starting with one vertex, and then at each step, we join a new vertex via one new edge to an existing vertex. This, in particular, implies that each tree graph has a leaf, i.e, a vertex of degree one. An example of a tree graph is given in Figure \ref{tree}.

\subsection{Equilibrium condition}
In this part, we show that if the graph $\mathbb{G}$ is a tree graph, then the set of equilibria associated with system \eqref{MODEL} can be characterized by a simple condition stated below. 
\vspace{5pt}

\begin{lem}\label{EQUIC}
Let $\mathbb{G}=(V,E)$ be a tree graph, then $p$ is an equilibrium associated with system \eqref{MODEL} if and only if $u_{ij}=0$ for all $(i,j)\in E$.  
\end{lem} 
\vspace{5pt}

\begin{proof}
It is clear that if each $u_{ij}$ is zero, then $p$ is an equilibrium. We now prove the other direction, and the proof is done by induction on the number of agents. 
\vspace{3pt}
\\
{\it Base case}. Suppose $N=2$, then $\dot{\vec a}_1= -\dot{\vec a}_2 = u_{12}\cdot (\vec a_1-\vec a_2)$. So if $p = (\vec a_1,\vec a_2)$ is an equilibrium, then either $u_{12}= 0$ or $\vec a_1=\vec a_2 = 0$, but they both imply  $u_{12}=0$.   
\vspace{3pt}
\\
{\it Inductive step}. Assume the lemma holds for $N\le k-1$ with $k\ge 3$, and we prove for the case $N=k$. Since each tree graph has at least one leaf, we may assume that vertex $k$ is a leaf of $\mathbb{G}$ and it joins the graph via edge $(1, k)$. Suppose $p = (\vec a_1,\cdots,\vec a_k)$ is an equilibrium, then we must have
\begin{equation}
\dot{\vec a}_k = u_{1k} \cdot(\vec a_k-\vec a_1) = 0  
\end{equation}
Then by the same arguments we used for proving $u_{12} = 0$ in the base case, we conclude that  
$
u_{1k} = 0
$. 
Now let 
\begin{equation}
V':= \{1,\cdots, k-1\}
\end{equation} 
and let 
$\mathbb{G}'=(V',E')$ be the subgraph induced by $V'$, i.e, for any two vertices $i$ and $j$ in $V'$, the pair $(i, j)$  is an edge of $\mathbb{G}'$ if and only if it is an edge of $\mathbb{G}$. It is clear that $\mathbb{G}'$ is also a tree graph.  Let $p'$ be a sub-configuration of $p$ consisting of agents $\vec a_1,\cdots,\vec a_{k-1}$, then $p'$ is also an equilibrium under $u_{ij}$ with $(i,j)\in E'$. This holds  because $p$ is an equilibrium and meanwhile $u_{1k}  = 0$, so the agent $\vec x_k$ doesn't attract or repel any agent in $p'$. By induction, we have 
$
u_{ij} = 0
$  
for all $(i,j)\in E'$. This then, combined with the condition $u_{1k} = 0$, establishes the proof.  
\end{proof}

\subsection{Geometry of the set of equilibria}
In this part, we will use the equilibrium condition to characterize the set of equilibria of system \eqref{MODEL}.
\vspace{5pt}

\begin{theorem}\label{CONTE}
Let $\mathbb{G}$ be a tree graph. Let $K$ be the set of equilibria of system \eqref{MODEL} with $u_{ij}$ defined by expression \eqref{DEFIU}. Then there is a diffeomorphism of $K$ given by
\begin{equation}
K \approx \Pi^{N-1}_{i=1}S^{n-1}
\end{equation}
as a product of $(N-1)$ copies of the unit sphere $S^{n-1}$ in $\mathbb{R}^{n}$.
\end{theorem}
\vspace{5pt}

\begin{proof}
The proof of the theorem will again carried out  by induction on the number of agents. So we first prove for the case $N=2$, and the inductive step will be given after that. 
\vspace{5pt}
\\
{\it Base case}. We show that Theorem \ref{CONTE} holds in the case $N=2$. Suppose $p = (\vec a_1,\vec a_2)$ is an equilibrium, then by Lemma \ref{EQUIC}, we have $u_{12}=0$, this then implies 
\begin{equation}\label{INNER}
\langle \vec b_1 -\vec b_2, \vec a_1-\vec a_2\rangle = | \vec a_1-\vec a_2 |^2
\end{equation} 
The set of equilibria associated with system \eqref{MODEL} is characterized by equation \eqref{INNER}, together with the condition that $\vec a_1+\vec a_2=0$.  Let $W$ be a subset of $\mathbb{R}^n$ defined by 
\begin{equation}\label{DEFW}
W:= \big \{\vec v\in \mathbb{R}^n\big | \langle \vec b_1 -\vec b_2, \vec v\rangle =  |\vec v|^2\big \}
\end{equation} 
Then it is clear that $K$ is diffeomorphic to $W$. To see this, we define a map $\varphi: W\rightarrow K$ by 
\begin{equation}
\varphi:\vec v\mapsto \frac{1}{2}(\vec v,-\vec v) 
\end{equation} 
It is clear that the map $\varphi$ is a diffeomorphism. We will now show that the set $W$ is itself a sphere in $\mathbb{R}^n$. Let 
\begin{equation}
\left\{
\begin{array}{ll}
\vec b:= \frac{1}{2}(\vec b_1-\vec b_2) \vspace{3pt}\\ 
 r :=\frac{1}{2} |\vec b_1-\vec b_2|
 \end{array}
 \right. 
 \end{equation}
and let $S_{r}(\vec b)$ be the sphere of radius $r$ centered at $\vec b$ in $\mathbb{R}^{n}$, i.e,  
\begin{equation}
S_{r}(\vec b):=\big  \{\vec v\in\mathbb{R}^n\big | |\vec v-\vec b|=r\big \}
\end{equation} 
It is clear by computation that 
\begin{equation}\label{Sbw}
S_r(\vec b) = W
\end{equation}  
In fact, if $\vec v\in\mathbb{R}^n$ lies inside $S_r(\vec b)$, then 
\begin{equation}\label{dayu}
\langle \vec b_1 -\vec b_2, \vec v\rangle > |\vec v|^2
\end{equation}
and if $\vec v$ lies outside  $S_r(\vec b)$, then
\begin{equation}\label{xiaoyu} 
\langle \vec b_1 -\vec b_2, \vec v\rangle < | \vec v|^2
\end{equation} 
This then completes the proof of the base case. 
\vspace{5pt}
\\
{\it Inductive step}. We will now use induction to prove Theorem \ref{CONTE}. We assume that the theorem holds for $N\le k-1$ with $k\ge 3$, and we prove for the case $N=k$. We again assume that vertex $k$ is a leaf of $\mathbb{G}$ and it joins the graph via edge $(1,k)$. Let $\mathbb{G}'=(V',E')$  be the subgraph of $\mathbb{G}$ induced by vertices $V'=\{1,\cdots,k-1\}$, then $\mathbb{G}'$ is a tree graph. Let $p'$ be a sub-configuration of $p$ consisting of agents $\vec a_1,\cdots,\vec a_{k-1}$.  Let $K'$ be a subset of $\mathbb{R}^{n\times(k-1)}$ defined by
\begin{equation}
K':=\big \{p'= (\vec a_1,\cdots,\vec a_{k-1})\big | p' \text{ is an equilibrium}\big \}
\end{equation} 
The equilibria set $K$ is then characterized by the condition that $p'$ is an equilibrium, together with the condition that $u_{1k} = 0$. Since these two conditions are independent of each other, there is a diffeomorphism of $K$ given by
\begin{equation}
K\approx K'\times S^{n-1}
\end{equation}
We may translate each $p'\in K'$ in $\mathbb{R}^n$ so that the centroid of $p'$ is zero after translation. Since $\mathbb{G}'$ is a tree graph,  by induction the set $K'$ is diffeomorphic to $\Pi^{k-2}_{i=1}S^{n-1}$, and hence $K$ is diffeomorphic to $\Pi^{k-1}_{i=1}S^{n-1}$.
\end{proof}
\vspace{5pt}

One may ponder at this point whether the existence of continuum of equilibria is a consequence of the fact that a tree graph is not a rigid graph. However, it is not the case.  For example, if we consider three agents evolving on a plane with $\mathbb{G}$ being the complete graph,  one can then show that the set of equilibria is diffeomorphic to a disjoint union of two circles. Though at this moment we do not have a statement  about the set of equilibria in the most general case, the tree-graph cases, as well as the  three-agents example suggest that it may be inevitable for system \eqref{MODEL} to possess a continuum of equilibria. 

\section{Simulated Annealing on Formation Control}\label{SA}
In the previous section, we have showed that there may exist a continuum of equilibria of system \eqref{MODEL} under the proposed formation control law. This  certainly affects the efficacy of the algorithm because there may exist a solution of system \eqref{MODEL} which converges to an equilibrium other than the target configuration.  
In this section, we will focus on fixing the problem. In view of the fact that there is only one local (global) minimum of the quadratic Lyapunov function $\Phi$ which is the target configuration, we attempt to apply simulated annealing, as a heuristic method, to the decentralized formation control  algorithm. In particular, if we add an appropriate noise term to each $u_{ij}$, the resulting stochastic system is then described  by
\begin{equation}\label{SGS}
d\vec a_i = \sum_{j\in V_i}(u_{ij}dt + \lambda_{ij}(t) dw_{ij}(t)) \cdot (\vec a_i-\vec a_j),  \hspace{10pt} \forall i = 1,\cdots,N
\end{equation}  
where the $w_{ij}(t)$ are independent standard Wiener processes, and $\lambda_{ij}(t)$ is a scalar function of time $t$ and $|\vec a_j-\vec a_i|$ defined by 
\begin{equation}\label{LAMBDA}
\lambda_{ij}(t) := 
\left\{
\begin{array}{ll}
c_1\exp(-c_2 t)/|\vec a_i-\vec a_j| & \text{if }\vec a_i\neq \vec a_j\vspace{3pt} \\
0 & \text{otherwise}
\end{array}\right.
\end{equation}
with $c_1$ and $c_2$  positive constants. As $\exp(-c_2 t)$ decays along time, the impact of the noise tends to zero as $t$ goes to infinity. With these noise terms, we expect that the centroid of the stochastic formation control system is still invariant along the evolution because otherwise  
the entire configuration may drift to some place  which is neither predictable nor controllable. Fortunately this is the case here as stated in the  next theorem.  
\vspace{5pt}

\begin{theorem} The centroid of the configuration is invariant along the evolution of the stochastic formation control system described by expression \eqref{SGS}.  
\end{theorem}
\vspace{5pt} 

\begin{proof}  Let $\phi:\mathbb{R}^N\rightarrow \mathbb{R}$ be a function defined by  
\begin{equation}
\phi(z_1,\cdots,z_N) := \sum^N_{i=1}z_i
\end{equation} 
Let $x^k_i$ be the $k$-th coordinate of agent $\vec a_i$. We need to show that  
\begin{equation}
d\phi(x^k_1,\cdots,x^k_N)=0
\end{equation} 
for all $k=1,\cdots, n$. Let $\vec x^k\in \mathbb{R}^N$ be a vector collecting the $k$-th coordinates of all the agents, i.e,   
\begin{equation}
\vec x^k := ( x^k_1,\cdots,x^k_N)
\end{equation}  
By defining vector $\vec x^k$, we can rewrite the system equation  \eqref{SGS} in a matrix form as 
\begin{equation}
d\vec x^k = U\vec x^kdt + \sum_{(i,j)\in E}\Lambda_{ij}(t)\vec x^kdw_{ij}  
\end{equation} 
where $U$ is a symmetric matrix of zero-column/row-sum with the $ij$-th, $i\neq j$, entry defined by
\begin{equation}
U_{ij} := \left\{
\begin{array}{ll}
-u_{ij} & \text{if } (i,j)\in E\vspace{3pt} \\
0 & \text{otherwise}
\end{array}\right.
\end{equation} 
and each $\Lambda_{ij}(t)$ is also a symmetric matrix of zero-column/row-sum defined by
\begin{equation}
\Lambda_{ij}(t) = \lambda_{ij}(t)\cdot (\vec e_i\vec e^T_i + \vec e_i\vec e^T_i-\vec e_i\vec e^T_j-\vec e_j\vec e^T_i)
\end{equation} 
where $\vec e_1,\cdots,\vec e_N$ is the standard basis of $\mathbb{R}^N$. 
\vspace{3pt}

We now apply the It\=o rule, and get the stochastic differential equation for $\phi(\vec x^k)$ as follows
\begin{equation}\label{ITO}
\begin{array}{lll}
d\phi(\vec x^k) & = & \Big \langle \frac{\partial \phi(\vec z)}{\partial \vec z}\Big |_{z = \vec x^k}, U\vec x^k \Big\rangle dt \vspace{5pt}\\ 
& & 
+ \frac{1}{2}\sum_{(i,j)\in E}\Big \langle \Lambda_{ij}\vec x^k, \frac{\partial^2 \phi(z)}{\partial \vec z^2}\Big |_{\vec z = \vec x^k}\Lambda_{ij}\vec x^k \Big \rangle dt \vspace{5pt}\\
 & & +\sum_{(i,j)\in E}\Big \langle  \frac{\partial \phi(z)}{\partial \vec z},\Lambda_{ij}\vec x^k\Big \rangle dw_{ij}
\end{array}
\end{equation}
Notice that for any $\vec z\in \mathbb{R}^N$, we have
\begin{equation}
\left\{
\begin{array}{l}
\frac{\partial \phi(\vec z)}{\partial \vec z} = \vec e\vspace{3pt}\\
\frac{\partial^2 \phi(z)}{\partial \vec z^2} = 0
\end{array}
\right.
\end{equation}
where $\vec e\in \mathbb{R}^N$ is a vector of all ones, and
\begin{equation}
\left\{
\begin{array}{l}
U \vec e = 0 \vspace{3pt}\\
\Lambda_{ij} \vec e = 0 \hspace{10pt} \forall (i,j)\in E
\end{array}
\right.
\end{equation}
So then all inner-products in equation   \eqref{ITO} vanish. This completes the proof. 
\end{proof}
\vspace{5pt}

\begin{figure}[h]
\includegraphics[scale=.47,left]{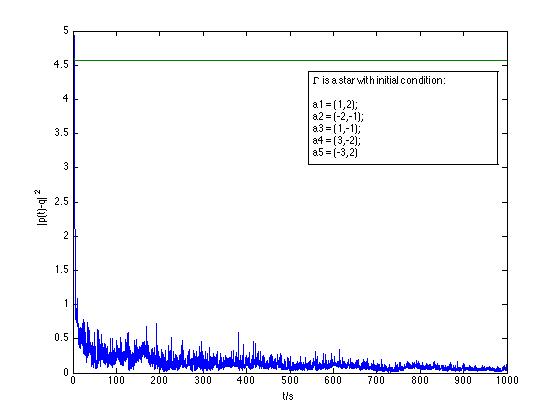}
\caption{This figure shows how $|p(t)-q|^2$ evolves over time $t$ with/without noise term under the condition that $\mathbb{G}$ is a star graph, with the initial condition given by expression \eqref{INIT1}.}
\label{SPStar}
\end{figure}

\begin{figure}[h]
\begin{center}
\includegraphics[scale=.47]{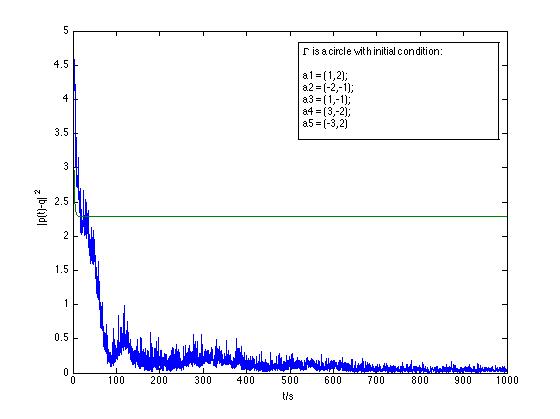}
\caption{This figure shows how $|p(t)-q|^2$ evolves over time $t$ with/without noise term under the condition that $\mathbb{G}$ is a circle, with the initial condition given by expression \eqref{INIT1}.}
\label{SPCircle}
\end{center}
\end{figure}

We  now give some examples of this stochastic formation control system, and illustrate how sample paths of this stochastic system evolve over time $t$. 
\vspace{5pt}
\\
{\bf Examples}. Consider five agents  $\vec a_1$, $\vec a_2$, $\vec a_3$, $\vec a_4$ and $\vec a_5$ evolving in $\mathbb{R}^2$. Let $q = (\vec b1,\vec b_2,\vec b_3,\vec b_4,\vec b_5)$ be the target configuration given by 
\begin{equation}
\begin{array}{lll}
\vec b_1 = (0,0) & \vec b_2 = (-1,1) & \vec b_3 = (1,1) \vspace{3pt} \\
 \vec b_4 = (1,-1) & \vec b_5 = (-1,-1) &
\end{array}
\end{equation}
We will work with two network topologies, one is a star graph which is a special type of tree graph and the other is  a circle. Details are described below  
\vspace{5pt}
\\
{\it 1. Star as the network topology}. We assume that $\mathbb{G}=(V,E)$ is a star graph with $E$ defined by
\begin{equation}\label{STAR}
E = \{(1,2),(1,3),(1,4),(1,5)\}
\end{equation}
We then pick an initial condition $p(0)$ given by 
\begin{equation}\label{INIT1}
\begin{array}{lll}
\vec a_1(0) = (1,2) & \vec a_2(0) = (-2,-1) & \vec a_3(0) = (1,-1)  \vspace{3pt}\\
 \vec a_4(0) = (3,-2) & \vec a_5(0) = (-3,2) &
\end{array}
\end{equation}
In Figure \ref{SPStar}, we show how the value of the quadratic Lyapunov function $\Phi(p(t))=|p(t)-q|^2$ evolves over time. The smooth curve (the green one) refers to the solution of system \eqref{MODEL} where there is no noise term added into the control law. As this solution converges  to an equilibrium, so we see from the figure that the green curve converges  to  a constant line along the evolution. Also it is clear that the solution, with the initial condition given by equation \eqref{INIT1}, does not converge to the target configuration. On the other hand, the ragged curve refers to the solution of system \eqref{SGS} where we have added noise terms into it. In the simulation, we have chosen $c_1 = 0.5$ and $c_2 = 0.001$. We see from the figure that $\Phi(p(t))$ approaches zero, in a stochastic way, along time $t$  which  implies that $p(t)$ approaches $q$.  
\vspace{3pt}  
\\  
{\it 2. Circle as the network topology}. We assume that $\mathbb{G}= (V,E)$ is now a circle with $E$ defined by
\begin{equation}\label{CIRC}
E = \{(1,2),(2,3),(3,4),(4,5),(1,5)\}
\end{equation}
We adopt the same initial condition given by expression \eqref{INIT1}. Figure \ref{SPCircle} shows how $\Phi(p(t))$ evolves over time $t$. Similarly, we see that if there is no noise term, then the solution of system \eqref{MODEL} does not converge to the target configuration. On the other hand, the sample path $p(t)$  approaches $q$ along the evolution. The two parameters $c_1$ and $c_2$  are again chosen to be $ 0.5$ and $ 0.001$ respectively.

\vspace{3pt}

These two examples have  demonstrated that  simulated annealing can be used to modify   the formation control law in order to achieve global convergence to the target configuration. More provable facts are needed at this moment for this heuristic algorithm.

\section{Conclusion}
In this paper, we have proposed a decentralized formation control law for agents to converge to a target configuration in the physical space. In particular, there is a quadratic Lyapunov function associated with the formation control system whose unique local (also global) minimum point is the target configuration.  
We then focussed on one of the limitations of this formation control model, i.e, there may exist a continuum of equilibria of the system, and thus there are solutions of system \eqref{MODEL} which do not converge to the target configuration. To fix this problem, we then applied the technique of simulated annealing to the formation control law, and showed that the modified stochastic system preserves one of the basic properties of the undirected formation control system, i.e, the centroid of the configuration is invariant along the evolution over time. We then worked on two simple examples of the stochastic system. Simulations results showed that sample paths approach to the target configuration.

\section{Acknowledgements}

The author here thanks Prof. Ali Belabbas, Prof. Tamer Ba\c sar, as well as the reviewers of the earlier draft of this work for their comments on this paper.


\begin{thebibliography}{99}

\bibitem{OR2} Olfati-Saber, R., \& Murray, R. M. (2002, July). ``Distributed cooperative control of multiple vehicle formations using structural potential functions." In IFAC World Congress (pp. 346-352).

\bibitem{GP}  Gazi, V., \& Passino, K. M. (2004). ``A class of attractions/repulsion functions for stable swarm aggregations." International Journal of Control, 77(18), 1567-1579.

\bibitem{KMB} Krick, L., Broucke, M. E., \& Francis, B. A. (2009). ``Stabilisation of infinitesimally rigid formations of multi-robot networks." International Journal of Control, 82(3), 423-439.


\bibitem{AB3} Belabbas, A., Mou, S., Morse, A. S., \& Anderson, B. D. (2012, December). ``Robustness issues with undirected formations." In CDC (pp. 1445-1450).

\bibitem{AH} Helmke, U., \& Anderson, B. D. (2013, October). ``Equivariant Morse theory and formation control." In Communication, Control, and Computing (Allerton), 2013 51st Annual Allerton Conference on (pp. 1576-1583). IEEE.

\bibitem{A3} Anderson, B. D., \& Helmke, U. (2014). ``Counting critical formations on a line." SIAM Journal on Control and Optimization, 52(1), 219-242.

\bibitem{USZB} Helmke, U., Mou, S., Sun, Z., \& Anderson, B. D. O. (2014). ``Geometrical methods for mismatched formation control." In Conference on Decision and Control.

\bibitem{ZSB} Sun, Z., Mou, S., Anderson, B. D., \& Morse, A. S. (2014). ``Formation movements in minimally rigid formation control with mismatched mutual distances." In The 53rd IEEE Conference on Decision and Control (CDC 2014).  


\bibitem{SAB} Mou, S., Morse, A. S., \& Anderson, B. D. O. (2014). ``Toward Robust Control of Minimally Rigid Undirected Formations." In The 53rd IEEE Conference on Decision and Control (CDC 2014).   


\bibitem{GFOM} Chen, X. (2014, June). ``Gradient flows for organizing multi-agent system." In American Control Conference (ACC), 2014 (pp. 5109-5114). IEEE.



\bibitem{CDC2014} Chen, X., \& Brockett, R. W. (2014). ``Centralized and Decentralized Formation Control With Controllable Interaction Laws." In The 53rd IEEE Conference on Decision and Control (CDC 2014).

\bibitem{CDC2015} Chen, X. ``Reciprocal Multi-Agent Systems with Triangulated Laman Graphs", preprint available on arXiv. 











\end{thebibliography}
\end{document}